\documentclass{article}

\usepackage[T1]{fontenc}
\usepackage[utf8]{inputenc}
\usepackage{microtype}
\usepackage{graphicx}
\usepackage{booktabs}
\usepackage{amsmath,amssymb,mathtools}
\usepackage{stmaryrd}
\usepackage{xspace}
\usepackage[dvipsnames]{xcolor}
\usepackage{multirow,makecell,array,tabularx}
\usepackage{subcaption}
\usepackage{xurl}
\usepackage{tikz}
\usepackage{eurosym}
\usetikzlibrary{positioning,arrows.meta,fit,calc,shapes.geometric,backgrounds}
\usepackage{hyperref}
\usepackage[accepted]{mlsys2025}
\makeatletter
\renewcommand{\Notice@String}{Preprint, November 2026.}
\makeatother

\newcommand{\fao}{\textsc{KathDB-FAO}\xspace}

\newcounter{examplectr}
\newcommand{\examplequery}[2][]{%
  \refstepcounter{examplectr}\par\smallskip\noindent
  \textbf{Example \theexamplectr.} \emph{#2}\ifx&#1&\else\label{#1}\fi\par\smallskip}

\hypersetup{pdfauthor={Guorui Xiao, Douglas Brown, Artur Borycki, Magdalena Balazinska},pdfsubject={arXiv preprint}}
\mlsystitlerunning{KathDB-FAO: Operator Grouping and Reuse}

\begin{document}

\twocolumn[
\mlsystitle{\fao: Optimizing Multimodal Queries through Operator Grouping and Code Reuse}
\begin{mlsysauthorlist}
\mlsysauthor{Guorui Xiao}{uw}
\mlsysauthor{Douglas Brown}{teradata}
\mlsysauthor{Artur Borycki}{teradata}
\mlsysauthor{Magdalena Balazinska}{uw}
\end{mlsysauthorlist}
\mlsysaffiliation{uw}{University of Washington}
\mlsysaffiliation{teradata}{Teradata}
\mlsyscorrespondingauthor{Guorui Xiao}{grxiao@cs.washington.edu}
\vskip 0.3in]
\printAffiliationsAndNotice{}

\begin{abstract}
Advances in multimodal large language models (multimodal LLMs) allow data systems to answer analytical queries over text and images alongside relational tables, but at a significant cost for large-scale data.
Existing systems reduce this cost through rewrite rules written in advance, which cannot use the logic of an individual query or the semantics of the data it processes.
To address this challenge, we design, implement, and evaluate \fao, a query evaluation system over multimodal data.
\fao decomposes a complex query into an execution plan of fine-grained steps, which we call Function-as-Operators (FAOs).
To reduce execution cost, it searches for subsets of these FAOs to implement together as one function, enabling query-specific optimizations that make fewer model calls.
Synthesizing these functions with a frontier model has its own cost, yet we observe that synthesized FAOs often recur across queries.
To further reduce plan generation cost, \fao uses pre-built functions and reuses functions synthesized for earlier queries.
On SemBench and an additional workload, \fao achieves quality comparable to the strongest competing system while reducing end-to-end cost per query by 76\% on average. 
\end{abstract}

\section{Introduction}\label{sec:intro}

\begin{figure}[t]
\centering
\includegraphics[width=\linewidth]{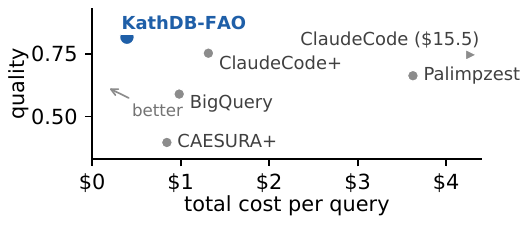}
\caption{Average total cost per query (planning plus execution) and quality across the three SemBench scenarios, weighting each scenario equally.
Table~\ref{tab:e2e} gives the detailed numbers.}
\label{fig:cost-quality}
\end{figure}

Analytical queries increasingly combine relational data with text and images, which multimodal LLMs such as GPT-4o can interpret~\cite{openai2024gpt4ocard}.
Interpreting such data can require a model call per record, which makes cost the central systems problem: processing 15\,GB of text with GPT-4o, for example, costs about \$18K~\cite{liu2025llmrelational}.
Consider a user with the following natural-language query:

\examplequery[ex:beach]{We are making a photo book of our past trips from our social-media posts and want one page of beaches. Show me beach photos from three different places we have been to.}

The first option for answering this query is to use a data system that extends SQL with semantic operators, which call models to perform inference on data~\cite{google2026bigqueryai,snowflake2025aisql}, or one that provides a programming interface for such operators, such as Palimpzest and LOTUS~\cite{liu2025palimpzest,patel2025lotus}.
These systems optimize queries by selecting among alternative model and operator implementations or applying rewrite rules~\cite{russo2026abacus,shankar2025moar}.
However, two limitations remain.
First, these optimizers rely on predefined operator implementations or rewrite directives~\cite{russo2026abacus,shankar2025moar}, limiting their ability to apply query-specific optimizations not covered by these predefined rules.
Second, users must express their queries in SQL or code, which can be a barrier for non-experts~\cite{gao2024dailsql}.

The second option is a natural-language query system such as CAESURA~\cite{urban2024caesura}, or a general coding agent such as Claude Code~\cite{anthropic2025claudecode}.
These systems construct query-specific plans or code from natural-language requests.
However, they lack dedicated optimizers that search for cheaper semantic query plans, and planning and code generation incur model costs for each query~\cite{bai2026agentcost}.
For our comparison, we use ClaudeCode$^+$, a variant of Claude Code that can call GPT-4o-mini for data processing to reduce execution cost.

\begin{figure}[t]
\centering
\includegraphics[width=\linewidth]{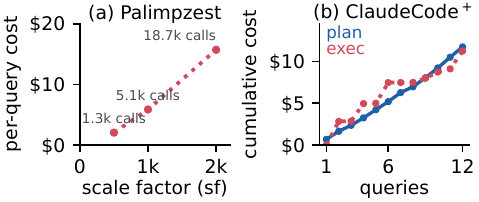}
\caption{Baseline costs on expensive queries over text and images; solid: planning, dotted: execution. (a) Palimpzest's execution cost per query grows with the data. (b) ClaudeCode$^+$'s cumulative costs grow with every query.}
\label{fig:cost-regimes}
\vspace{-0.2cm}
\end{figure}

Figure~\ref{fig:cost-regimes} illustrates the two costs: Palimpzest's execution cost grows with the data size, while ClaudeCode$^+$ incurs planning costs for each query, so its cumulative planning cost grows across queries.

To reduce execution cost, our key insight is that carefully choosing which relational and semantic steps to implement together enables query-specific optimizations that can avoid unnecessary model calls.
In Example~\ref{ex:beach}, each post has one caption but may contain many photos.
After extracting place names from captions, a straightforward plan classifies every photo before selecting beach photos from three different places.
A better plan jointly implements photo classification and selection, stopping after finding the first beach photo for each place.
An even more efficient plan leverages the extracted place names when generating the joint function, allowing it to prioritize locations more likely to have beaches (e.g., Waikiki before Zermatt) and stop once three places qualify.
At the other extreme, generating the entire query as one function remains challenging for complex tasks~\cite{le2024codechain}.
It also prevents the generated code from exploiting optimizations that depend on observing intermediate values, such as the extracted place names, which are not yet available when the function is generated.
Traditional operator fusion often reduces intermediate materialization and data movement~\cite{chen2018tvm,neumann2011efficiently}; here, the combined function also eliminates model calls using the query's logic and the intermediate data.

Additionally, we observe that synthesized functions can often be reused across queries, since queries in a workload tend to repeat~\cite{vanrenen2024redshift} or share parts of their plans.
For example, if the user later asks for beach photos from five places rather than three, adapting the earlier function may require only a small change instead of generating a new function from scratch.
Reusing generated functions can therefore reduce planning cost.

In this paper, we design, implement, and evaluate
\fao, a query
evaluation subsystem for FAO-based multimodal DBMSs that reduces execution
cost through query-specific grouping and planning cost through code reuse.

Given a natural-language query, it decomposes the query into fine-grained steps, selects groups for joint implementation, reuses previously generated functions when possible, and executes the resulting plan.

Building \fao raises two challenges.
First, choosing which steps to implement together requires comparing the expected execution costs of candidate groupings.
Traditional cost-based optimizers estimate the costs of alternative plans~\cite{selinger1979access}.
Here, however, the potential savings depend on query-specific optimizations in code that has not yet been generated.
Generating and testing every candidate would make planning prohibitively expensive.
Second, reuse must be both inexpensive and faithful to the current query.
Finding and adapting saved code incurs overhead, while using code from a similar but logically different query without adaptation can produce incorrect results.

To address the first challenge, \fao profiles individual steps on a small data sample and compares candidate groupings using query logic, plan structure, and observed intermediate results, without estimating numerical execution costs or synthesizing every candidate.
To address the second challenge, \fao identifies reusable fused implementations by matching their constituent steps and inputs against parts of the current query's plan, and adapts saved code to the current logic only when the corresponding group is selected.
For common steps, \fao selects and configures pre-built functions rather than synthesizing new code, reducing the code-generation cost to operator selection and argument specification.
\fao optimizes the execution plan and is complementary to techniques that lower the cost of each call~\cite{kwon2023vllm,zheng2024sglang,chen2023frugalgpt,liu2025llmrelational}.

In summary, this paper makes the following contributions:

\begin{itemize}

\item \textbf{Query-specific grouping and synthesis (\S\ref{sec:optimizer})}:
\fao searches for groups of relational and semantic steps that can be jointly implemented to eliminate unnecessary model calls, without relying on predefined rewrite rules.
It selects promising groupings without synthesizing every candidate and generates their implementations after their inputs are materialized, enabling optimizations that exploit query logic and intermediate data.

\item \textbf{Cross-query reuse of synthesized optimizations (\S\ref{sec:reuse})}:
\fao reuses synthesized fused FAOs across queries by matching them to subplans of later queries and adapting their code to the new query logic.
It further reduces planning cost through pre-built operators for atomic FAOs, even when fused-code reuse is unavailable.

\item \textbf{System implementation and evaluation (\S\ref{sec:arch}, \S\ref{sec:eval})}:
We implement \fao and show that it achieves comparable quality while reducing end-to-end cost per query by 76\% on average relative to the highest-quality external system in each workload (Table~\ref{tab:e2e}).

\end{itemize}

\section{System Overview and Problem Formulation}
\label{sec:arch}\label{sec:system}

\begin{figure}[t]
\centering
\includegraphics[width=\linewidth]{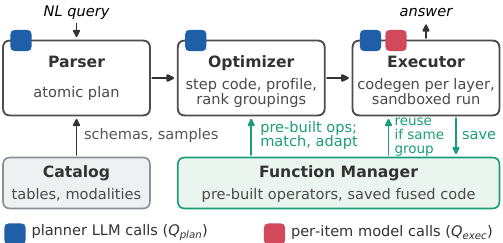}
\caption{Architecture of \fao.}
\label{fig:system}
\end{figure}

\fao answers natural-language queries over tables that contain relational data and paths to image and text files. It does so by building and executing query plans (Figure~\ref{fig:system}).
The Parser breaks a query into an atomic plan of steps.
The Optimizer profiles these steps and chooses which to implement together to reduce model calls (\S\ref{sec:optimizer}).
The Executor generates or reuses code for the selected groups, executes the plan, and saves code for later queries (\S\ref{sec:reuse}).

\subsection{From Natural Language to an Atomic Plan}
\label{sec:setup}

\textbf{Parser.}
Given a natural-language query $q$ over a multimodal database $D$, the Parser prompts an LLM with the query, each table's schema (column names, data types, primary key, and modality annotations), and three sample rows per table, following prior work~\cite{gao2024dailsql,rajkumar2022evaluating,urban2024caesura}.
The LLM then decomposes the query into fine-grained steps that form an \emph{atomic plan}: a directed acyclic graph (DAG) in which each node represents a step in natural language and each edge represents a data dependency.
Here, ``atomic'' refers to the initial, ungrouped steps produced by the Parser rather than a unique minimum decomposition.
Translating natural-language queries into correct atomic plans is a separate research problem~\cite{li2023bird,lei2025spider2}.
While \fao generates these plans, this paper focuses on optimizing a given atomic plan rather than studying NL-to-plan correctness.
Thus we assume that the atomic plan correctly captures the query, and users can inspect or update it before execution.

\fao represents each step as a \emph{Function-as-Operator} (FAO), separating what the step computes from how the system will implement it.

\noindent\textbf{Definition 1 (Function-as-Operator).}
An FAO is a tuple
\[
f=\langle \mathit{ID},\tau,\mathit{desc},\mathit{in},\mathit{out},\mathit{impl}\rangle,
\]
where $\tau$ is the step's operator type (relational or semantic), $\mathit{desc}$ describes its logic in natural language, and $\mathit{in}$ and $\mathit{out}$ specify its input and output relations.
The implementation $\mathit{impl}$ is initially empty and may later be filled with pre-built, newly generated, or adapted code.
A RELATIONAL FAO does not call models, and its type is an operator of the extended relational algebra~\cite{codd1970relational,garciamolina2008database} (e.g., $\mathsf{REL\text{-}JOIN}$).
A SEMANTIC FAO calls a model to interpret text or images, as in semantic operators~\cite{patel2025lotus}.
The Parser returns an atomic plan $S=(F,E)$, where $F$ is the set of FAOs and $E$ represents their data dependencies.

\subsection{From Atomic Plans to Execution}

\textbf{Plans and grouping.}
The atomic plan specifies the query's logical steps and data dependencies, while grouping determines which steps are jointly implemented.
A \emph{plan} $P$ partitions $F$ into groups, each implemented as one FAO $f_g$ that takes one or more relations as input and produces one or more relations as output.
We call $f_g$ \emph{fused} when its group contains two or more FAOs; otherwise, the group consists of a single atomic FAO.
Within a fused FAO, semantic operations and their downstream relational steps can share control flow, allowing downstream conditions to determine when further model calls are unnecessary.

We denote by $\mathcal P_K(S)$ the set of candidate plans whose groups contain at most $K$ FAOs and form a DAG.
Because our objective is to reduce model-call costs, we restrict fused groups to contain at least one SEMANTIC FAO.
It must also be \emph{convex}, meaning every data-flow path between its members stays within the group.
This prevents a data dependency from leaving a group and returning to it, which could create a cycle.

\begin{figure}[t]
\centering
\includegraphics[width=\linewidth]{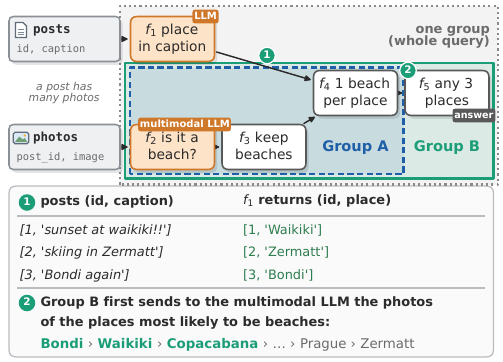}
\caption{Atomic plan of Example~\ref{ex:beach}. Orange FAOs call models to interpret captions or photos; white FAOs are RELATIONAL.}
\label{fig:atomic}\label{fig:plan}\label{fig:rewrite}
\vspace{-0.5cm}
\end{figure}

Figure~\ref{fig:plan} shows the atomic plan for Example~\ref{ex:beach}.
Here, $f_1$ extracts place names from post captions, $f_2$ classifies photos, and $f_3$--$f_5$ select beach photos from three different places.
Group~A jointly implements $f_2$--$f_4$ to stop classifying photos after finding the first beach photo for each place.
Group~B additionally includes $f_5$, allowing it to stop once three places qualify.

\textbf{Contracts.}
Since code is generated independently for each FAO, an FAO may omit columns needed by downstream steps.
For example, in Figure~\ref{fig:plan}, $f_1$ extracts place names from captions, but $f_4$ also needs the post ID to associate each place with its photos.
If $f_1$ returns only place names, $f_4$ cannot join its output with \texttt{photos.post\_id}.
To prevent such mismatches, \fao derives \emph{contracts} before code generation.
An LLM derives these contracts by propagating output requirements \textit{backward} from the query's requested output through the atomic plan, specifying the required columns, types, and expected values.

\textbf{Optimizer.}
The Optimizer must select a grouping before its fused implementations exist, making their execution costs difficult to estimate. To avoid generating and running the code for each possible grouping, the Optimizer
profiles atomic FAOs on sample data, recording intermediate results and model-call statistics.
Using these profiles and the query logic and plan structure, it uses an LLM-based ranking algorithm to compare candidate groupings by their relative expected execution costs, and select a promising plan $P\in\mathcal P_K(S)$ without synthesizing the candidates.

\textbf{Executor.}
The Executor processes the selected groups in dependency order.
For each fused group, it adapts saved code whose steps and inputs match the group when possible, or synthesizes a new implementation otherwise.
The code is generated or adapted after the group's input relations have been materialized, allowing it to use actual intermediate values when determining which model calls to make.

\subsection{Objectives}
\label{subsec:cost}

As in other semantic data systems~\cite{shankar2025moar,russo2026abacus}, we focus on the monetary cost of model calls as the metric to minimize.
In our evaluated workloads, local pandas operations are much faster than model inference calls.
We thus exclude local CPU and I/O costs from our monetary cost model.

\textbf{Cost model.}
For a model call $c$, let $t_{\mathrm{in}}(c)$ and $t_{\mathrm{out}}(c)$ denote its billed input and output tokens, and let $p_{\mathrm{in}}(c)$ and $p_{\mathrm{out}}(c)$ denote the corresponding prices in USD per token.
The call costs $\mathrm{cost}(c)=p_{\mathrm{in}}(c)t_{\mathrm{in}}(c)+p_{\mathrm{out}}(c)t_{\mathrm{out}}(c)$ USD.

We consider a workload $W=(q_1,\ldots,q_n)$ of consecutive queries over $D$.
Each query incurs planning and execution costs related to model calls at those steps:
\begin{equation}\label{eq:workload}
Q(W)
=\sum_{i=1}^{n}
\big(q_i^{\mathrm{plan}}+q_i^{\mathrm{exec}}\big)
=Q_{\mathrm{plan}}+Q_{\mathrm{exec}}.
\end{equation}

For query $q_i$ with selected plan $P_i$, we decompose these costs as
\begin{equation}\label{eq:plan}
\begin{aligned}
q_i^{\mathrm{plan}}
&=q_i^{\mathrm{parse}}
+q_i^{\mathrm{prof}}
+q_i^{\mathrm{rank}}
+q_i^{\mathrm{syn}},\\
q_i^{\mathrm{exec}}
&=\sum_{g\in P_i}
\sum_{c\in\mathcal C(f_g;D)}
\mathrm{cost}(c),
\end{aligned}
\end{equation}
where $q^{\mathrm{parse}}$ covers constructing the atomic plan and propagating contracts, $q^{\mathrm{prof}}$ covers implementing atomic FAOs and profiling them on samples, $q^{\mathrm{rank}}$ covers ranking candidate groupings, and $q^{\mathrm{syn}}$ covers generating or adapting fused implementations.
Each planning component sums the monetary costs of model calls made during the corresponding stage.
The multiset $\mathcal C(f_g;D)$ contains model calls made by group $g$ when processing $D$, including repeated calls.


Grouping targets execution cost, while pre-built operators and code reuse target planning cost.
\fao addresses these two objectives separately rather than jointly minimizing total workload cost, and it does not guarantee optimality.

\textbf{Execution cost (\S\ref{sec:optimizer}).}
Given an atomic plan $S_i$, the objective is to find the grouping with the lowest execution cost:
\begin{equation}\label{eq:problem}
P_i^*
=\arg\min_{P\in\mathcal P_K(S_i)}
q_i^{\mathrm{exec}}(P).
\end{equation}

\textbf{Planning cost (\S\ref{sec:reuse}).}
Let $L_{i-1}$ denote the fused FAOs saved from earlier queries.
\fao reduces planning cost in two ways.
First, it uses pre-built operators to avoid generating code for atomic FAOs during profiling.
Second, after selecting a grouping $P_i$, it adapts matching fused code from $L_{i-1}$ instead of synthesizing new implementations.
Together, these techniques aim to reduce $q_i^{\mathrm{prof}}+q_i^{\mathrm{syn}}$.

Jointly optimizing planning, model execution, and conventional query-processing costs, such as CPU and I/O, is left to future work.

\section{Execution Cost Reduction}\label{sec:optimizer}

The goal of execution cost reduction is to select the lowest-cost plan from $\mathcal P_K(S)$ (Equation~\ref{eq:problem}).
Here, a critical challenge is that a grouping's savings depend on the fused code that implements it and the data it processes, making those savings difficult to estimate without actually generating and executing that code.
We address this challenge by comparing candidate groupings using atomic-plan information and sampled execution profiles, synthesizing fused code only for the selected plan (Algorithm~\ref{alg:loop}).

\begin{algorithm}[t]
\caption{Searching \& executing a low-cost query plan}
\label{alg:loop}
\small
\begin{algorithmic}[1]
\REQUIRE atomic plan $S$, data $D$, cap $K$, width $w$
\STATE $H\gets\textsc{Profile}(S,D)$\label{line:profile}
\STATE $C\gets\mathcal P_K(S)$\label{line:enum}
\WHILE{$|C|>1$}\label{line:rank-start}
  \STATE $C\gets\textsc{ListRank}(C,H,w)$\label{line:rank}
\ENDWHILE
\STATE $P\gets$ the plan left in $C$\label{line:pick}
\FOR{each group $g\in P$ in topological order}\label{line:loop}
  \IF{$|g|=1$}
    \STATE $f_g\gets$ the code of $g$ in $H$
  \ELSE
    \STATE $f_g\gets\textsc{ReuseOrSynthesize}(g,H,\text{inputs of }g)$\label{line:syn}
  \ENDIF
  \IF{$f_g$ fails after three attempts}
    \STATE $f_g\gets$ the code of $g$'s FAOs in $H$
  \ENDIF
  \STATE run $f_g$\label{line:run}
\ENDFOR
\end{algorithmic}
\end{algorithm}

\emph{Profiling and candidates (lines~\ref{line:profile} and~\ref{line:enum})}:
To give the ranker evidence about the data, \fao executes the atomic FAOs on samples of the base tables, using pre-built operators or newly generated code.
These runs produce a profile $H$ containing their code, intermediate results, selectivities, model-call counts, and model responses. In our evaluation, the sample has at most 50 rows per base table; Appendix~A varies the sample size.
\fao then enumerates the candidate plans in $\mathcal P_K(S)$ without implementing them.
In our benchmarks, the most complex query has 156 candidates, making enumeration practical at this scale; larger plans could use search methods such as iterative dynamic programming~\cite{kossmann2000iterative}.

\emph{Plan selection (lines~\ref{line:rank-start}--\ref{line:pick})}:
Traditional cost-based query optimizers estimate numerical execution costs for candidate plans and select the one with the lowest estimated cost~\cite{selinger1979access}.
Applying this approach to synthesized FAOs is challenging because their implementations do not yet exist, and their model-call counts can depend on data-dependent branches and early termination.
We initially explored two estimation strategies: \emph{llm-guess}, which synthesizes each candidate and asks an LLM to predict its model-call count, and \emph{stub-fn}, which executes synthesized candidates on a sample with stubbed model calls and extrapolates the count.
We observed that \emph{llm-guess} could produce inconsistent numerical estimates across candidates, while \emph{stub-fn} could miss branches that eliminate model calls on the full dataset.
Both approaches also require synthesizing every candidate, incurring substantial planning overhead.

However, the atomic plan already specifies each step's query-specific logic and dependencies, revealing potential optimizations such as stopping model calls once a downstream limit is satisfied.
Profiling further provides observed selectivities and intermediate values that can help assess the potential savings.
These observations motivate comparing candidate plans based on their potential savings, rather than predicting a numerical execution cost for each one.

We therefore use \emph{list-rank}, a tournament-style LLM ranking method inspired by LLM-as-a-judge~\cite{zheng2023judging}.
Each call compares up to $w$ candidate plans using the original NL query, the atomic FAO descriptions and profiled code, and sample statistics such as row counts, selectivities, and frequent values.
The LLM ranks candidates by expected execution cost, considering which model calls could be avoided through joint implementation and the tokens required per call.
For the top-ranked plan, it also explains the potential optimizations, which the Code Generator receives as a hint when synthesizing its fused FAOs.
We evaluate \emph{list-rank} against both estimation-based alternatives in \S\ref{subsec:eval-grouping}.

When there are more than $w$ candidates, \fao ranks them in batches and advances each batch's winner to the next round until one plan remains.
For a fixed $w>1$, the number of ranking calls grows linearly with the number of enumerated candidates; with $w=20$, ranking 156 candidates takes 9 calls.

\begin{figure}[t]
\centering
\includegraphics[width=\linewidth]{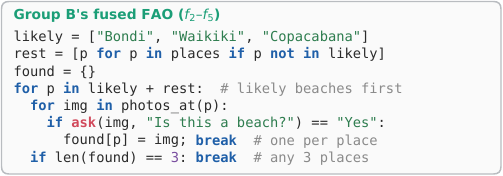}
\caption{Synthesized code of Group~B's FAO. The function prioritizes places likely to contain beaches, stops at each place's first beach photo, and terminates after finding three qualifying places.}
\label{fig:fused}
\vspace{-0.5cm}
\end{figure}

\emph{Synthesis and fallback (lines~\ref{line:loop}--\ref{line:run})}:
Once a plan is selected, the Executor runs its groups in topological order.
Singleton groups use the atomic code already stored in $H$.
For each fused group, \textsc{ReuseOrSynthesize} adapts saved code when its atomic steps and inputs match, or synthesizes a new function otherwise (\S\ref{sec:reuse}).
For newly synthesized fused FAOs, code generation occurs after their input relations have been materialized, allowing the Code Generator to use their schemas and sampled values alongside the atomic implementations and the ranker's optimization hint.
Figure~\ref{fig:fused} shows this: Group~B uses the place names returned by $f_1$ to prioritize likely beach locations and jointly implements photo classification with downstream stopping conditions.
If the code fails after three attempts, the Executor runs the group's atomic FAOs instead. We did not observe this fallback in our evaluation.

\section{Plan Generation Cost Reduction}
\label{sec:executor}\label{sec:reuse}

\fao incurs code generation costs at two stages: implementing atomic FAOs for profiling ($q^{\mathrm{prof}}$) and synthesizing fused FAOs after grouping ($q^{\mathrm{syn}}$).
To reduce these costs, \fao uses pre-built operators for atomic FAOs and reuses fused code generated for earlier queries.

\textbf{Pre-built atomic FAOs.}
We observe that many atomic FAOs perform common operations, so generating new code for each one is often unnecessary.
\fao therefore provides thirteen pre-built operators.
Six are semantic operators in the style of LOTUS~\cite{patel2025lotus} (e.g., a semantic filter that uses an LLM to decide whether a condition holds for a record).
The other seven are relational operators (e.g., a relational filter that evaluates a condition on column values without model calls).

During profiling, \fao prompts an LLM whether a pre-built operator can implement an atomic FAO, using its description, input columns and sample values, and output contracts.
If a suitable operator exists, the LLM selects it and fills in its arguments (e.g., join keys).
\fao then validates the proposed call \textit{without an LLM} (e.g., referenced columns exist in the inputs).
If the checks pass, \fao runs the operator on the sample; otherwise, or if execution fails, it synthesizes new code.
Here, using a pre-built operator can reduce both input and output tokens: the prompt omits instructions required for free-form code generation (e.g., the \texttt{Available Libraries} list specifying the libraries available), and the LLM only selects an operator and specifies its arguments rather than generating an entire function.

\textbf{Reusing fused FAOs.}
Queries in a workload often share parts of their plans, but similar plan structures do not necessarily imply the same computation.
For example, the fused function in Figure~\ref{fig:fused} prioritizes likely beach locations and stops after finding one beach photo from each of three places.
A later query asking for two sunset beach photos from each of five places could have the same operator types and dependencies but require different classification and stopping conditions, and potentially a different order in which to examine locations.
Directly executing the saved code would produce incorrect results, while generating a new function would discard potentially reusable optimizations such as early stopping.
Thus, \fao adapts saved fused code to the current query logic while retaining applicable optimizations.

After answering a query, \fao saves each executed fused FAO in the Function Manager.
Each entry stores the fused code and metadata about its atomic FAOs (e.g., operator types, descriptions).
Saving requires no additional LLM calls.
The library holds at most $N$ fused FAOs ($N=20$ in our evaluation) and evicts the least recently used entry when full.

When a new query arrives, \fao identifies and retains potential reuse candidates by matching saved fused FAOs against parts of the atomic plan using operator types, data dependencies, and input tables.
After the Optimizer selects a plan (\S\ref{sec:optimizer}), \fao checks whether any retained candidate matches a selected fused group's atomic FAOs and inputs.
If a matching candidate is available, \fao attempts to adapt its saved code; otherwise, it synthesizes a new implementation.

For each selected match, \fao asks an LLM to adapt the saved fused code to implement the corresponding group of FAOs in the current plan.
The LLM receives the saved code, the descriptions and output contracts of the saved and current atomic FAOs, and the current group's inputs and atomic implementations.
It makes targeted edits (e.g., changing beach to sunset beach) while keeping applicable parts and logic of the saved implementation, avoiding regeneration of the entire function.
\fao then validates the edited code without an LLM.
It rejects code that does not parse, and if this adaptation is rejected, \fao synthesizes a new fused implementation; otherwise, the Executor runs the adapted code.
\S\ref{subsec:eval-reuse} evaluates the resulting planning-cost savings, including the overhead of reuse and pre-built operators. 

\section{Evaluation}\label{sec:eval}

\begin{table*}[t]
\centering\small
\setlength{\tabcolsep}{2pt}
\caption{End-to-end quality and cost (USD per query);
best in bold, second best underlined.
$^*$: at least one 12-hour timeout, scored 0. OOM: out of memory. N/A: planning included in execution (ClaudeCode) or plans written by experts. Palimpzest and BigQuery run SemBench's expert-written plans, so they have no planning cost and we do not run them on Additional, which would need new hand-written plans or queries.
Comparisons use unrounded values.}
\label{tab:e2e}
\begin{tabular}{l ccc ccc ccc ccc}
\toprule
& \multicolumn{3}{c}{Additional} & \multicolumn{3}{c}{MMQA (sf200)} & \multicolumn{3}{c}{E-Com (sf500)} & \multicolumn{3}{c}{Movie (sf2000)} \\
\cmidrule(lr){2-4}\cmidrule(lr){5-7}\cmidrule(lr){8-10}\cmidrule(lr){11-13}
System & quality$\uparrow$ & plan \$$\downarrow$ & exec \$$\downarrow$ & quality$\uparrow$ & plan \$$\downarrow$ & exec \$$\downarrow$ & quality$\uparrow$ & plan \$$\downarrow$ & exec \$$\downarrow$ & quality$\uparrow$ & plan \$$\downarrow$ & exec \$$\downarrow$ \\
\midrule
CAESURA$^+$    & OOM & OOM & OOM & 0.53 & 0.89 & 0.154 & 0.20 & 0.83 & 0.174 & 0.46 & 0.48 & \textbf{0.003} \\
ClaudeCode     & 0.86 & N/A & 4.399 & 0.70 & N/A & 34.676 & 0.72 & N/A & 10.520 & \textbf{0.82} & N/A & 1.402 \\
ClaudeCode$^+$ & 0.92 & 1.33 & 0.201 & 0.69 & 1.49 & 0.065 & 0.79 & 1.19 & 0.176 & 0.78 & 1.01 & 0.007 \\
\midrule
Palimpzest     & -- & -- & -- & 0.78$^*$ & N/A & 7.161 & 0.66 & N/A & 2.402 & 0.55$^*$ & N/A & 1.309 \\
BigQuery       & -- & -- & -- & 0.32 & N/A & 0.142 & 0.70 & N/A & 2.254 & 0.75 & N/A & 0.553 \\
\midrule
\shortstack[l]{\fao\\[-2pt]{\scriptsize atomic}}
& \textbf{0.94} & \underline{0.54} & 0.127
& \textbf{0.88} & \textbf{0.28} & 0.028
& \textbf{0.82} & \underline{0.54} & 0.187
& \underline{0.81} & \underline{0.33} & 0.009 \\
\shortstack[l]{\fao\\[-2pt]{\scriptsize grouping}}
& 0.92 & 0.66 & \underline{0.014}
& \underline{0.86} & 0.38 & \textbf{0.026}
& \underline{0.80} & 0.60 & \textbf{0.109}
& 0.77 & 0.43 & \underline{0.003} \\
\shortstack[l]{\fao\\[-2pt]{\scriptsize grouping + reuse \& pre-built}}
& \underline{0.93} & \textbf{0.43} & \textbf{0.013}
& 0.86 & \underline{0.28} & \underline{0.027}
& 0.79 & \textbf{0.47} & \underline{0.117}
& 0.80 & \textbf{0.28} & 0.004 \\
\bottomrule
\end{tabular}
\end{table*}

We answer three questions:
(RQ1)~How do \fao's end-to-end quality and cost compare with existing systems (\S\ref{subsec:e2e})?
(RQ2)~How much execution cost does grouping save, and can alternative plan-selection methods, operator reordering, or predefined grouping rules achieve similar savings (\S\ref{subsec:eval-grouping})?
(RQ3)~How much planning cost does reuse save for workloads of similar queries over the same data (best case) versus diverse queries over different data (worst case) (\S\ref{subsec:eval-reuse})?
Appendix~B studies impact of using alternative models.

\subsection{Experimental Setup}\label{subsec:setup}\label{subsec:impl}

We use three SemBench scenarios~\cite{lao2025sembench}: \emph{Movie} analyzes the sentiment of movie reviews~\cite{andrezaza2023clapper} (sf2000), \emph{E-Com} queries fashion products with text and images~\cite{aggarwal2019fashion} (sf500), and \emph{MMQA} answers questions over tables, text, and images~\cite{talmor2021mmqa} (sf200).
The scale factor (sf) specifies the number of rows in each scenario's primary entity table, with the sizes of other tables determined by sf.
A larger sf increases the dataset size and generally leads to more model calls.
We leave out the two scenarios with audio queries, because current models understand these audio recordings poorly~\cite{lao2025sembench}.
We exclude two queries from E-Com and one query from MMQA because some baselines do not support them~\cite{lao2025sembench}.
\fao itself supports every query as it generates code for each FAO, so it is not limited to a fixed set of operators, which is one advantage of generating code over composing operators.
We further use an \emph{Additional} workload that has ten harder queries over the same data.

Following SemBench~\cite{lao2025sembench}, we measure quality using query-specific metrics that compare each query's result with its ground truth.
Specifically, it is F1 for retrieval, accuracy for classification, the Adjusted Rand Index for grouping, $\max(0,\rho)$ for ranking, and $1/(1+\text{relative error})$ for aggregation.
Costs are USD per query from token counts and list prices: planning cost is $q^{\mathrm{plan}}$ and execution cost is $q^{\mathrm{exec}}$ of Equation~(\ref{eq:plan}).
We average 3 runs for each query and then over a scenario's queries.
A run that exceeds 12 hours scores 0 and keeps its cost.
We report cost rather than latency as our runs share nodes and an API proxy with other users of a multi-tenant cluster, and model API latency varies with load, so wall-clock time has a high variance.

We compare against five systems.
CAESURA$^+$~\cite{urban2024caesura} (using claude-opus-4-7 as planner) translates a query into relational and modality-specific operators, with LLM-based operators in place of its pretrained models.
ClaudeCode~\cite{anthropic2025claudecode} (using claude-opus-4-7) writes and runs Python end to end, and ClaudeCode$^+$ may also call gpt-4o-mini; since it interleaves planning with execution, we report ClaudeCode's total cost as execution cost.
For ClaudeCode$^+$, the Claude tokens count as planning and the gpt-4o-mini tokens as execution.
Palimpzest~\cite{liu2025palimpzest} runs SemBench's expert-written plans with gpt-4o-mini over data, and BigQuery~\cite{google2026bigqueryai} numbers are SemBench's published results with Gemini-2.5-Flash~\cite{gemini2025}.

The Parser, ranker, and Code Generator use claude-opus-4-7, and FAOs call gpt-4o-mini.
We use $K=5$, $w=20$, and a 50-row profiling sample (Appendix~A evaluates alternatives).
Each run uses two CPU cores and 128\,GB of memory.

\subsection{End-to-End Results (RQ1)}\label{subsec:e2e}

\begin{figure}[t]
\centering
\includegraphics[width=\linewidth]{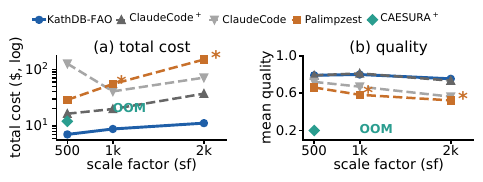}
\caption{E-Com: total cost over the queries (a) and mean quality (b) at scale factors 500, 1000, and 2000 (rows in the primary entity table). $*$: at least one query reached the 12-hour cap.}
\label{fig:scalability}
\vspace{-0.5cm}
\end{figure}

Figure~\ref{fig:cost-quality} plots total cost (planning + execution) per query against quality, averaged across the three SemBench scenarios.
\fao sits in the top-left corner: it has the highest quality and the lowest total cost.
Table~\ref{tab:e2e} breaks this down by scenario.
\fao achieves higher quality than all external systems on Additional and MMQA, matches the best on E-Com, and is within 0.02 of the best on Movie.
Compared with the highest-quality external system in each workload, \fao reduces end-to-end cost per query by 76\% on average.

We further evaluate \fao's scalability on SemBench's E-Com scenario, focusing on a single scenario to keep experimental costs manageable.
Figure~\ref{fig:scalability} shows how costs grow with data size.
\fao achieves the lowest total cost at every scale.
Palimpzest's cost grows super-linearly because its semantic join makes a model call for each pair of rows.
It reaches the 12-hour limit on two queries at sf1000 and three at sf2000.
CAESURA$^+$ runs out of memory above sf500, even with 128\,GB of RAM.

\subsection{Grouping (RQ2)}
\label{subsec:eval-grouping}\label{subsec:eval-selection}

\textbf{Comparing plan-selection methods.}
We compare plan-selection methods over the same candidate space $\mathcal P_K(S)$, using the same atomic plan and profiling results.
\emph{atomic} executes the atomic plan without grouping and serves as our baseline.
\emph{list-rank} is \fao's selection method (\S\ref{sec:optimizer}).
A natural alternative is to estimate each candidate's execution cost and select the cheapest~\cite{selinger1979access}.
To do so we evaluate the following alternatives:
\emph{llm-guess} synthesizes each candidate group's code and asks an LLM to estimate its full-data model-call count using the code and atomic profile results.
\emph{stub-fn} executes each candidate plan on the profiling sample, replacing model calls with stubbed responses.
It builds a response distribution for each model-call site from the atomic plan's profiling results, samples from these distributions when evaluating alternative groupings, and linearly extrapolates the observed call counts to the full dataset.
Both use dynamic programming to select the lowest-estimated-cost grouping and regenerate the selected fused code for execution.
We also compare against operator reordering and rule-based grouping below.

\begin{figure}[t]
\centering
\includegraphics[width=\linewidth]{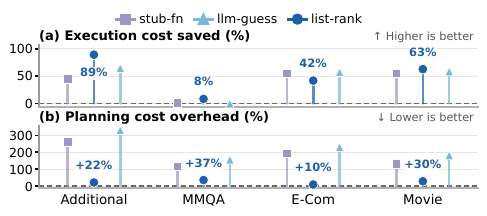}
\caption{Comparison of plan-selection methods across four workloads.
(a) Execution-cost savings (higher is better) and (b) planning-cost overhead
(lower is better), both relative to the atomic plan (dashed line at 0\%).
Stems show the differences from atomic, and blue labels show list-rank's values. 
Mean quality differs from the atomic plan by at most $0.05$ across methods and workloads and is not shown for space.}
\label{fig:rq2-quality}
\end{figure}

Figure~\ref{fig:rq2-quality} shows the results.
On Additional, \emph{list-rank} achieves the lowest execution cost at $0.11\times$ that of the atomic plan, outperforming \emph{llm-guess} and \emph{stub-fn}.
On E-Com and Movie, \emph{list-rank} also reduces execution cost to $0.58\times$ and $0.37\times$ that of the atomic plan, respectively.
On MMQA, many queries require a model response for \textit{every} record, and no method can reduce execution cost much.

For planning cost, \emph{list-rank} adds little overhead compared with the atomic plan.
In contrast, \emph{llm-guess} and \emph{stub-fn} incur $2.2$--$4.3\times$ the planning cost because they synthesize every candidate, yet neither achieves lower end-to-end cost than \emph{list-rank} on any workload.
On E-Com, these two methods' lower execution costs come from using inaccurate keyword matching rather than better grouping decisions (average F1 on two affected queries: 0.90 vs.\ 0.98 for \emph{list-rank}).

\begin{figure}[t]
\centering
\includegraphics[width=\linewidth]{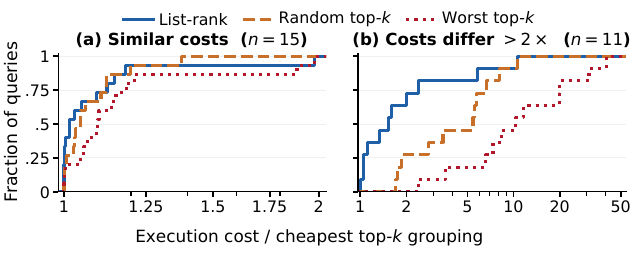}
\caption{Execution cost of list-rank's choice, a random selection, and the worst grouping, each normalized to the cheapest grouping among list-rank's top-$k$ candidates.
(a) Queries whose top-$k$ candidates differ in cost by at most $2\times$.
(b) Queries whose top-$k$ candidates differ by more than $2\times$.}
\label{fig:rankcheck}
\end{figure}

\textbf{Checking list-rank's choices.}
We also examine how good its selected plans are compared with other candidates.
We sample 20 queries from the four workloads and include all ten Additional queries (26 unique queries after deduplication).
For each query, we execute its top-$k$ candidates, where $k=\min(10,\text{number of candidates})$.

Figure~\ref{fig:rankcheck} separates the queries based on how much their top-$k$ candidates differ in execution cost.
For some queries, even the most and least expensive candidates are within $2\times$, leaving limited room for improvement.
When costs differ by more than $2\times$, list-rank substantially outperforms alternatives.

\textbf{Reordering and rule-based grouping.}
We next examine whether operator reordering or predefined grouping rules can achieve similar savings.
We focus on Additional, where grouping provides the largest execution-cost reduction.

\emph{reordering} mimics a classical database optimization for expensive predicates: evaluating them after cheaper filters~\cite{hellerstein1993predicate}.
Rather than implementing a reordering optimizer, we compute an optimistic estimate using full-data statistics.
We execute each atomic plan on the full dataset, recording the number of input and output rows and the cost per input row for every FAO.
We then enumerate legal operator orders and estimate their costs using these measurements.

Two FAOs may exchange positions if neither requires columns produced by the other.
A model call may also move across an inner join if it reads columns from only one side.
We additionally allow early termination for LIMIT without ORDER BY~\cite{graefe1993query,carey1997saying} and charge one model call per distinct input~\cite{hellerstein1996caching}.
This gives reordering access to full-data statistics rather than estimates from a sample.

\begin{figure}[t]
\centering
\includegraphics[width=\linewidth]{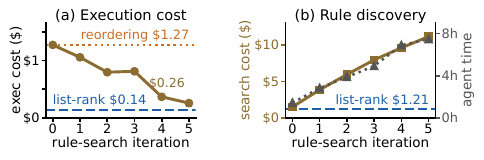}
\caption{Rule-based grouping on Additional.
(a) Total execution cost over the ten queries by iteration; line: rule-based; dotted: reordering; dashed: list-rank. (b) Cumulative rule-search cost (squares, left axis) and time (triangles, right axis); dashed: list-rank's \$1.21 extra planning cost over atomic across the ten queries.}
\label{fig:ladder}
\end{figure}

On Additional, reordering provides no savings (Figure~\ref{fig:ladder}).
The downstream filters that could reduce model calls depend on the outputs of preceding SEMANTIC FAOs, so moving them before those FAOs would change the query's computation.
For example, a filter on an LLM-generated label cannot execute before that label is produced.
Grouping instead allows these dependent operations to be implemented together, enabling optimizations such as early stopping.

\emph{rule-based} uses predefined grouping rules constructed by a Claude Code agent (claude-opus-5-5).
Each rule matches a pattern in the atomic plan, specifies which FAOs to group, and provides a fixed optimization hint to the Code Generator.
Rules may refer to operator types and data dependencies, but not to specific tables, columns, values, or queries.
Once constructed, applying the rules to select a grouping requires no LLM calls.
To be fair, we let the agent develop the rules iteratively for the Additional workload.
At each iteration, it examines the atomic plans, the fused code generated by list-rank, and the measured costs.
It then proposes one new rule and executes all ten queries with the accumulated rules.
The agent retains changes that reduce execution cost or restore quality and stops when it finds no further improvements.
For example, one rule groups a model call with the filters and limits that consume its output, allowing the fused code to stop making calls once a key's result is determined.

As Figure~\ref{fig:ladder} shows, the rules recover most of list-rank's execution-cost savings, but discovering them costs about nine times as much as list-rank's plan selection and code synthesis.
Additionally, these rules are fixed in advance and select groups based only on operator types and dependencies, without inspecting the actual input values.
For example, \fao can use materialized brand names to prioritize those likely to sell ethnic wear, reducing model calls in ways that the fixed rules cannot specify.

\textbf{Break-even analysis for \textit{list-rank}.}
Grouping reduces execution cost but adds planning cost, so we study when it becomes cheaper overall than atomic execution.
We use E-Com because its dataset is large enough to study this tradeoff as input size grows.
We evaluate all 12 E-Com queries and the 6 E-Com queries in Additional; we exclude the other four Additional queries because their MMQA dataset cannot be scaled to the same size.

To avoid expensive model calls at every scale, we replay the generated code using recorded model responses and costs, sampling responses from the same call site for unseen requests.
Planning cost remains fixed at its measured value.
To validate replay accuracy, we replay each run using half of its recorded model responses and compare the execution cost with the original run.
The difference averages 2\% and is at most 6.6\% across atomic and grouped plans here.

\begin{figure}[t]
\centering
\includegraphics[width=\linewidth]{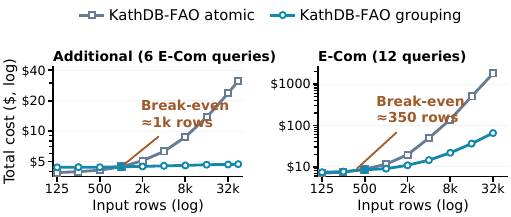}
\caption{Total query cost of atomic execution and list-rank grouping as the number of inputs grows (both axes logarithmic).}
\label{fig:grouping-scale}
\end{figure}

Figure~\ref{fig:grouping-scale} shows that grouping becomes more cost-effective as the dataset size grows.
However, grouping does not pay off on MMQA at the evaluated scale, as these queries require model calls for nearly every record, leaving little execution cost to save.

\subsection{Reuse (RQ3)}
\label{subsec:eval-reuse}\label{subsec:eval-fn-reuse}

This section evaluates how much \fao reduces planning cost by reusing fused FAOs and using pre-built atomic operators.
We construct two ten-query sequences to examine when these techniques are most and least effective.

In the \emph{best} sequence, queries ask the same question over the same data but vary in thresholds and filter conditions.
This reflects the repetition observed in real workloads: most queries in production warehouses repeat~\cite{vanrenen2024redshift}, and even exploratory SQL workloads contain repeated query-plan templates~\cite{jain2016sqlshare}.
In the \emph{worst} sequence, we randomly select queries from the four scenarios in Table~\ref{tab:e2e} (seed 100).

Each sequence starts with an empty library of capacity $N=20$.
\fao saves each executed fused FAO immediately and uses LRU eviction with a capacity of 20 entries, which neither sequence reaches.
We compare four settings.
\emph{no reuse} synthesizes both atomic and fused FAOs from scratch.
\emph{reuse} reuses saved fused FAOs when the Optimizer selects a matching group, adapting the saved code to the new query (\S\ref{sec:reuse}).
\emph{pre-built} uses one of the thirteen generic operators to implement an atomic FAO when applicable, reducing the cost of profiling (\S\ref{sec:reuse}).
\emph{pre-built + reuse} enables both techniques.

\begin{figure}[t]
\centering
\includegraphics[width=\linewidth]{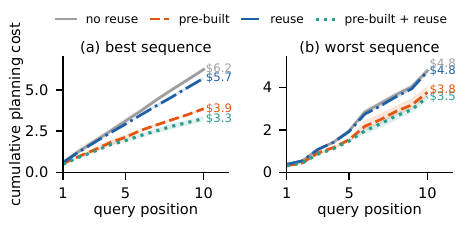}
\caption{Cumulative planning cost over the best (a) and worst (b) query sequences.
At position $i$, each line shows the total planning cost of the first $i$ queries.}
\label{fig:reuse}
\end{figure}

\begin{table}[t]
\centering\small
\setlength{\tabcolsep}{2pt}
\caption{Statistics for \emph{pre-built + reuse} on the two sequences (mean of three runs).
A saved FAO hits when its fused code is executed; a pre-built operator hits when it replaces an atomic FAO.}
\label{tab:reuse-stats}
\begin{tabular}{lcc}
\toprule
 & best & worst \\
\midrule
saved-FAO hit rate & 63\% & 0\% \\
pre-built FAO hit rate & 94\% & 92\% \\
pre-built lookup cost & \$0.58 & \$1.00 \\
planning cost saved, net of overhead & \$2.97 (48\%) & \$1.25 (26\%) \\
saved FAOs at the end (capacity 20) & 10.3 & 9.7 \\
\bottomrule
\end{tabular}
\end{table}

Figure~\ref{fig:reuse} and Table~\ref{tab:reuse-stats} show planning-cost savings of 48\% on the best sequence and 26\% on the worst, net of lookup overhead.
Fused-code reuse has a 63\% hit rate on similar queries but none on diverse queries.
In contrast, pre-built operators replace most atomic FAOs, reducing planning cost even without fused-code reuse.
The small cost difference between the two pre-built settings on the worst sequence reflects run-to-run variance.
Execution cost and quality remain largely unchanged.

\section{Related Work}\label{sec:rw}

\textbf{Semantic operators and expensive predicates.}
Semantic operator systems~\cite{patel2025lotus,liu2025palimpzest,shankar2025docetl,google2026bigqueryai,snowflake2025aisql} expose predefined operator interfaces, and their optimizers~\cite{russo2026abacus,satriani2025galois,shankar2025moar} search over alternative implementations, models, and rewrites.
KathDB proposed FAOs as an abstraction for multimodal queries~\cite{xiao2026kathdb}.
DocETL uses an LLM to apply predefined rewrite directives for text data~\cite{shankar2025docetl}, while CADENZA divides a semantic operator into subtasks handled by rules, smaller models, or LLMs, tuned using \textit{labeled} validation data~\cite{ha2026cadenza}.

Traditional database techniques reorder or cache expensive predicates~\cite{hellerstein1993predicate,hellerstein1996caching}, stop after enough results~\cite{carey1997saying}, or adapt operator ordering based on observed selectivity~\cite{avnur2000eddies}.
Proxy-model methods use cheaper models or indexes to filter and prioritize records~\cite{kang2017noscope,kang2019blazeit,kang2020supg,kang2022tasti}.
These approaches rely on predefined optimizations.
In contrast, \fao synthesizes code after a group's inputs are materialized, allowing it to prioritize records based on the meaning of their values without predefined rewrite rules or trained proxies.

\textbf{Fusion and code generation.}
Compilers fuse operators with fixed semantics to reduce execution overhead~\cite{chen2018tvm,neumann2011efficiently}.
MOAR~\cite{shankar2025moar} rewrites sequences of LLM operators over text documents and evaluates candidate accuracy using labeled samples or an LLM judge.
In contrast, \fao searches groupings of DAGs over tables, text, and images, ranking candidates by expected cost from one profiling run without labels.
Its fused FAOs decide at run time which model calls are necessary.
Evaporate synthesizes extraction functions from sample documents to approximate LLM outputs~\cite{arora2023evaporate}, rather than implementing a query plan.
Natural-language query systems and coding agents generate code or tool plans per query~\cite{trummer2022codexdb,urban2024caesura,nooralahzadeh2024multi,zhang2025vocaludf,anthropic2025claudecode}, but do not search for cheaper groupings.

\textbf{Reusing code.}
Multi-query optimization shares work among queries that run together~\cite{sellis1988multiple}, while parametric query optimization prepares plans for unknown run-time parameter values~\cite{ioannidis1992parametric}.
\fao instead reuses synthesized code across sequential queries.
Semantic caches reuse responses~\cite{bang2023gptcache}, tool libraries parameterize functions when saved~\cite{cai2024latm,yuan2024craft}, and library learning extracts reusable functions from programs~\cite{ellis2021dreamcoder,bowers2023stitch,grand2024lilo}.
\fao matches saved fused FAOs to the new atomic plan and adapts their code at reuse time, drawing inspiration from anti-unification~\cite{plotkin1970note}.

\textbf{Cheaper model calls.}
Serving systems and cascades reduce the latency or cost of individual calls~\cite{kwon2023vllm,zheng2024sglang,gim2024promptcache,liu2025llmrelational,chen2023frugalgpt}; \fao reduces the number of calls, making these approaches complementary.

\section{Conclusion}\label{sec:conclusion}

\fao answers natural-language queries over multimodal data.
It reduces execution cost by synthesizing groups of operators that exploit query logic and data semantics, avoiding unnecessary model calls without predefined rewrite rules.
Pre-built operators and cross-query code reuse reduce planning cost.
On SemBench and an additional workload, \fao achieves comparable quality at substantially lower end-to-end cost.

\bibliographystyle{mlsys2025}
\bibliography{ref}
\section*{AI Usage Disclosure}
We used Claude and ChatGPT to assist with language and grammar editing, and Claude Code to assist with writing figure-generation scripts and submitting batch jobs for experiments on a multi-tenant cluster.
\clearpage
\setcounter{figure}{0}
\renewcommand{\thefigure}{S\arabic{figure}}
\section*{Supplementary Material}
\appendix

\section{Knobs of the Optimizer}\label{app:knobs}\label{app:sample}

\begin{figure}[H]
\centering
\includegraphics[width=\linewidth]{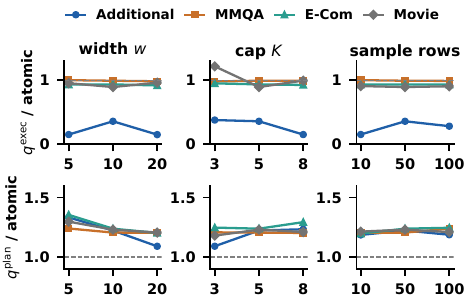}
\caption{List-rank with one knob varied at a time, on the four workloads; the other two knobs stay at the center of the sweep, $w=10$, $K=5$, and a 50-row profiling sample. Execution and planning cost relative to the atomic plan; one run per point.}
\label{fig:knobs}
\end{figure}

Figure~\ref{fig:knobs} varies the three knobs of \S\ref{sec:optimizer} of the main paper one at a time: the list width $w$, the group-size cap $K$, and the profiling sample.
Execution cost is largely stable on MMQA and E-Com, but varies more on Additional and, for small $K$, Movie.
Planning cost generally falls as $w$ grows, because a wider list takes fewer ranking rounds; changing $K$ or the profiling sample has a smaller effect on planning cost.
We use $w=20$, $K=5$, and 50 rows; $w=20$ is one point of the sweep.

\newpage
\section{Impact of other models}\label{app:eval-models}

\begin{figure}[H]
\centering
\includegraphics[width=\linewidth]{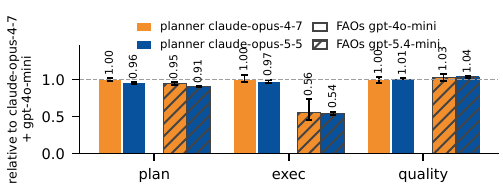}
\caption{\fao on Additional using different planning models (claude-opus-4-7 or claude-opus-5-5) and FAO models (gpt-4o-mini or gpt-5.4-mini), relative to the setting with claude-opus-4-7 and gpt-4o-mini.
Bars show the mean of three runs, and whiskers show the range.}
\label{fig:models}
\end{figure}

Figure~\ref{fig:models} evaluates \fao with different models for planning and FAO execution on Additional.
Switching the planning model from claude-opus-4-7 to claude-opus-5-5 has little effect on the results.
Using gpt-5.4-mini instead of gpt-4o-mini for FAO calls roughly halves execution cost while maintaining similar or slightly higher quality.
This is largely because most model calls on Additional process images, which cost approximately half as much with gpt-5.4-mini.

Across all model combinations, \fao reduces execution cost to $0.10$--$0.11\times$ that of the atomic plan using the same models.
Thus, \fao's execution-cost savings remain consistent across different planning and FAO models.

\end{document}